\documentclass[aps, prr, english, notitlepage, twocolumn, superscriptaddress, floatfix, longbibliography]{revtex4-2}
\usepackage{amssymb,amsmath,amsfonts,dsfont, amsthm}
\usepackage{comment}
\usepackage{graphicx}
\usepackage{dcolumn}
\usepackage{newtxtext,newtxmath,bm}
\usepackage{hyperref}
\usepackage{url}

\usepackage{color}

\begin{document}

\title{Fermionic tensor network contraction for arbitrary geometries}

\author{Yang Gao}%
\affiliation{Division of Engineering and Applied Science, California Institute of Technology, Pasadena, California 91125, USA}

\author{Huanchen Zhai}
\affiliation{Division of Chemistry and Chemical Engineering, California Institute of Technology, Pasadena, California 91125, USA}

\author{Johnnie Gray}
\affiliation{Division of Chemistry and Chemical Engineering, California Institute of Technology, Pasadena, California 91125, USA}

\author{Ruojing Peng}
\affiliation{Division of Chemistry and Chemical Engineering, California Institute of Technology, Pasadena, California 91125, USA}

\author{Gunhee Park}
\affiliation{Division of Engineering and Applied Science, California Institute of Technology, Pasadena, California 91125, USA}

\author{Wen-Yuan Liu}
\affiliation{Division of Chemistry and Chemical Engineering, California Institute of Technology, Pasadena, California 91125, USA}

\author{Eirik F.~Kj{\o}nstad}
\affiliation{Division of Chemistry and Chemical Engineering, California Institute of Technology, Pasadena, California 91125, USA}

\author{Garnet Kin-Lic Chan}
\email{gkc1000@gmail.com}
\affiliation{Division of Chemistry and Chemical Engineering, California Institute of Technology, Pasadena, California 91125, USA}

\date{\today}

\begin{abstract}
We describe our implementation of fermionic tensor network contraction on arbitrary lattices within both a globally ordered and locally ordered formalism. We provide a pedagogical description of these two conventions as implemented for the \texttt{quimb} library. Using hyperoptimized approximate contraction strategies, we present benchmark fermionic projected entangled pair states simulations of finite Hubbard models defined on the three-dimensional diamond lattice and random regular graphs.
\end{abstract}

\maketitle

\section{Introduction}

Tensor networks (TN) provide a compact representation of quantum states that have led to conceptual and computational advances across many areas of statistical mechanics, quantum science, and computer science~\cite{Orus2019}.
In the quantum setting, the basic idea is to represent the amplitudes of a many-body quantum wavefunction through the contraction of tensors arranged according to some network or graph. The first tensor network algorithm was the density matrix renormalization group (DMRG)~\cite{white1992density,white1993density}, which today is often viewed as an energy optimization technique for a class of tensor networks called matrix product states (MPS)~\cite{SCHOLLWOCK201196, CiracPerezSchuch2021}.
Since then, more general families of tensor networks  have been proposed, two common examples being the
projected entangled pair states (PEPS)~\cite{Murg2007, CiracPerezSchuch2021} and the multiscale entanglement renormalization ansatz (MERA)~\cite{vidal2007, vidal2008}.

In recent years, tensor networks have begun to be applied to complex geometries, including three-dimensional lattices and random graphs~\cite{gray2018quimb, Gray2021hyperoptimized, Gray2024hyperoptimized, ChenHelms2022, jahromiThermalBosons3d2020, vlaarSimulationThreedimensionalQuantum2021, PanZhouLiZhang2020, WangZhangPanZhang2024tn, sahu2022efficienttensornetworksimulation, patra2024projectedentangledpairstates}. For an exact tensor network contraction, the contraction order matters significantly for the overall computational cost. In an approximate tensor network contraction, the order of contraction further affects the accuracy. The \texttt{quimb} library discussed in the current work~\cite{gray2018quimb} was written to support tensor networks defined on arbitrary graphs and provides optimized heuristics for exact and approximate tensor network contraction on such graphs~\cite{Gray2021hyperoptimized, Gray2024hyperoptimized}.

Simulating many-body fermionic systems is a major application area for tensor networks. The most common representation of fermionic quantum states is via fermionic tensor networks, where the tensors may themselves be viewed as fermionic operators~\cite{kraus2010fermionic, Barthel2009fermionic, pineda2010fermionic}. As a result, changing the order of tensor contractions changes the value of the tensor network unless the fermionic signs are accounted for. In the earliest graphical implementations of fermionic tensor networks~\cite{corboz2009fmera, corboz1,corboz2, corboz2011, corboz2014, zheng2017stripe, benedikt2021scipost}, the contraction of the fermionic tensor network was performed as that of a bosonic tensor network with geometry-specific swap tensors inserted by hand. However, this strategy is not convenient for users of a general library, such as \texttt{quimb}, that is designed to support arbitrary tensor network graphs.

Here, we report on the fermionic tensor network formalism for arbitrary lattices and connectivities available through the \texttt{quimb} library.
While we have used these implementations already in prior publications (e.g. Refs.~\cite{Lee2023, liu2024tensor}), the methodology was not previously described.
In addition, there is more than one way to implement the fermionic tensor network formalism, and we describe the relationship between the ``globally ordered'' and ``locally ordered'' formalisms.
Below, we give a pedagogical introduction to fermionic tensor networks and the implementation of contraction on arbitrary lattices.
There is also another recent work that describes how to implement fermionic tensor network contraction in a lattice agnostic manner~\cite{ZHANG2024109355, mortier2024fermionic}, which uses the ``locally ordered'' formalism. Our presentation covers some similar ground but is complementary in other respects. As an illustration, we present initial benchmark studies on the ground-state of the Fermi-Hubbard model on the three-dimensional diamond lattice and on random regular graphs of degree 3.

\section{Fermionic tensor network computation}

\subsection{Standard tensor networks and symmetries}

In a standard (bosonic) tensor network contraction, the tensors are simple arrays of numbers. Each dimension of the array will be referred to as a leg or bond, and the contraction of two tensors corresponds to a partial summation of a common bond. For example, given two tensors $A$, $B$, we can perform a contraction as
\begin{align}
    C_{ijlmnr} = \sum_k A_{ijkl} B_{mnkr}. \label{eq:simplecontract}
\end{align}
A tensor network $\mathcal{T}$ arises from the contraction of an unordered set of tensors
\begin{align}
    \mathcal{T} = \mathcal{C} (A B C \ldots) \label{eq:tncontract}
\end{align}
where we use $\mathcal{C}$ to denote the contraction of all shared bonds.

It is common to consider symmetries in tensor networks~\cite{Singh2010, Singh2011, Singh2012}. For Abelian symmetries, we can use Abelian symmetric tensors in the tensor network~\cite{Singh2010, Singh2011}. The indices of Abelian symmetric tensors carry an additional integer irreducible representation (irrep) label, e.g., for $A_{ijkl}$, then
for an Abelian group of dimension $H$, each index $i,j,k,l$ has $H$ sectors for each irrep. For example, if $H=2$, then there are 4 sectors of a 2-leg tensor $A$,
\begin{align}
    A_{i_0j_0}, A_{i_0j_1}, A_{i_1j_0} A_{i_1j_1},
\end{align}
which we can write more succinctly as $A_{i_\sigma j_{\sigma'}}$ for irreps $ \sigma,\sigma' \in \{0, 1\}$. Writing the set $\{i \}$ as $\{ i_\sigma\}, \sigma=0\ldots H-1$, then we extract the irrep of a specific index as $s(i_\sigma)=\sigma$.

We will consider only symmetric tensors with well-defined total irrep. To define how the total irrep
is obtained from the irreps of the indices, we label each index of the tensor with a plus (+, ``outgoing'') or minus (-, ``incoming''). 
Then, for example,
if for $A_{ijkl}$ the labels of $ijkl$ are ${-}{+}{-}{+}$, we have
\begin{align}
A_{ijkl} = 0 ~~ \mathrm{if} ~~ -s(i)+s(j)-s(k)+s(l) \neq s(A) \label{eq:symmetryrule}
\end{align}
where $s(A)$ is the total irrep of $A$. 
Similarly if $ijkl$ are labelled ${+}{+}{+}{+}$, the entries of $A_{ijkl}$ vanish if $+s(i)+s(j)+s(k)+s(l) \neq s(A)$, and so on.
Note that for finite Abelian groups (e.g., $\mathbb{Z}_2$), addition and subtraction is modulo $H$.

Tensor decomposition routines are an integral part of approximate tensor network algorithms. The QR factorization and singular value decomposition (SVD) are examples of special interest, which are based on first treating the tensors as matrices by grouping their indices.
For standard tensors, this amounts to reordering the desired row and column indices to be adjacent so that $
    A_{ijkl} \to A_{ij,kl},
$
where $ij$ is now the row index and $kl$ is now the column index for the matrix decomposition.
In the case of symmetric tensors, there is a degree of freedom on how to partition the input total symmetry in the output, i.e., for the QR decomposition, $s(A)  = s(Q) + s(R)$, while for SVD, $s(A) = s(U) + s(S)  + s(V^\dag)$, where $S$ is the diagonal matrix consisting singular values.
If $s(A)$ is not the totally symmetric irrep, we can use the convention that only one of the decomposed tensors carries the irrep $s(A)$ and the others are totally symmetric. For example, in the QR decomposition,
\begin{align}
    A_{ij,kl} \to \sum_{mn} Q_{ij, mn} R_{mn,kl}
\end{align}
$Q$ has the same dimensions as $A$, and we choose $s(Q)=s(A)$, and the legs of $Q$ carry the same symmetry labels as the legs of $A$. In the SVD, we {choose $s(U)=s(A)$}, and the singular values are truncated up to dimension $\chi$ using the singular values from all the symmetry blocks of $S$.

\begin{figure}
    \centering
    \includegraphics[width=\linewidth]{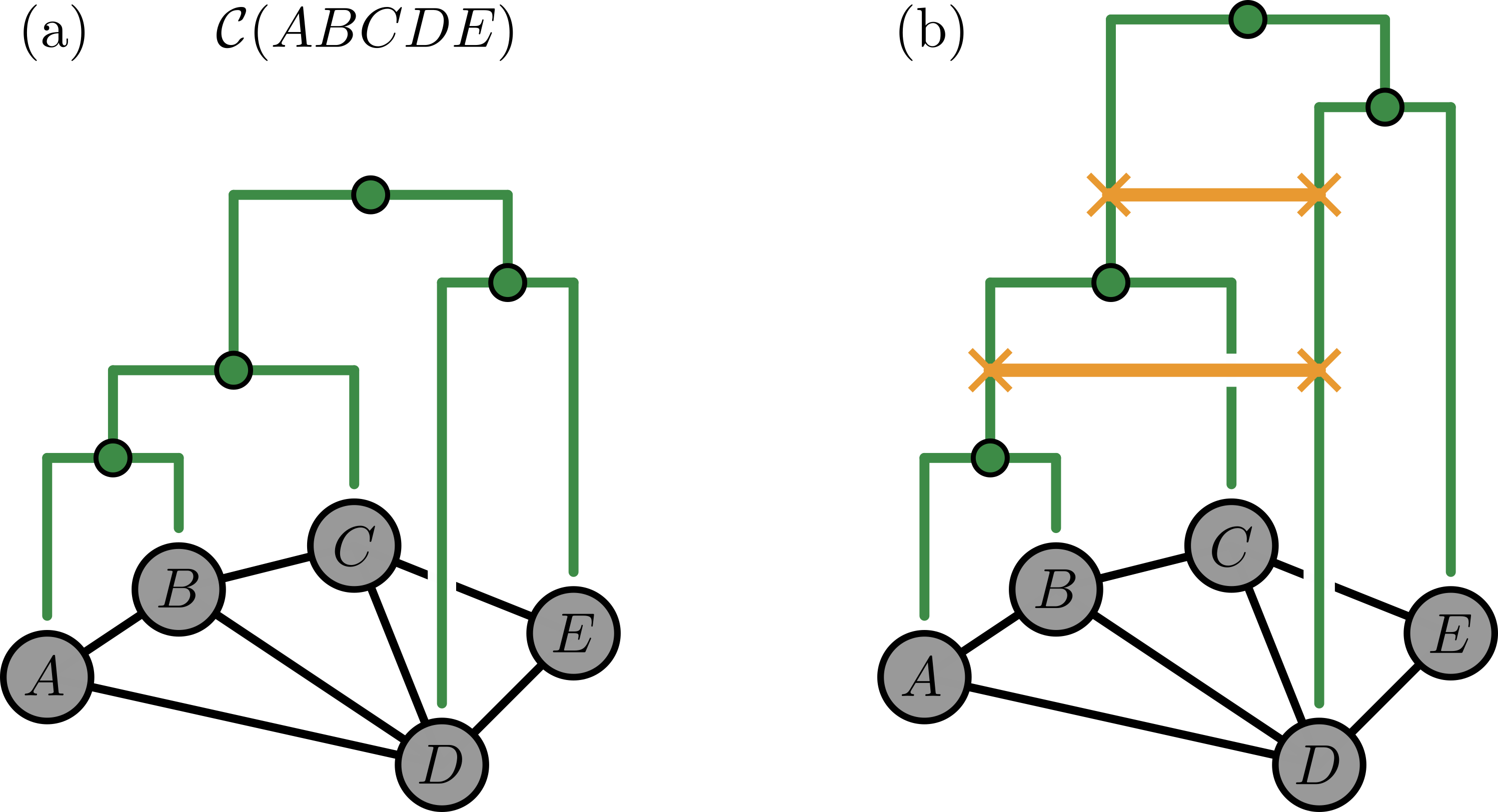}
    \caption{Examples of contracting a tensor network (a) exactly and (b) approximately.
    Computation flows from bottom to top.
    In each case, pairwise contractions form a tree (green lines), with intermediate tensors shown as smaller green circles.
    In the approximate case, compressions between intermediates also occur, shown here as horizontal orange lines connecting crosses.
    }
    \label{fig:contractiontree}
\end{figure}

The unordered contraction in Eq.~\ref{eq:tncontract} defines the value of $\mathcal{T}$. For symmetric tensors, the +/- labels of the contracted indices must be compatible for the contraction to be valid, that is, the outgoing leg of one tensor can only be contracted with the ingoing leg of another (in Eq.~\ref{eq:simplecontract}, either $k$ is labelled + on $A$ and - on $B$, or vice versa), and any such compatible choice of +/- labels is a valid symmetric tensor network $\mathcal{T}$. 
However, the cost to compute $\mathcal{T}$ further depends on the order of the tensor contractions. If approximate contraction is used, both cost and accuracy also depend on when and which shared bonds are compressed. The order of tensor contraction in the exact case defines a contraction tree (see Fig.~\ref{fig:contractiontree}(a)) where the vertical axis illustrates the time-ordering of the pairwise contractions. In the approximate contraction case, Fig.~\ref{fig:contractiontree}(b), the contraction tree contains additional lines between branches indicating the compression of bonds between tensors.

\subsection{Fermionic tensors and contractions}

\label{sec:fermionicten}

A fermionic tensor network is formed from fermionic tensors. We thus first discuss the properties of fermionic tensors. In our implementation, these are stored as Abelian arrays of numbers with additional metadata for parity and other symmetries, and the internal index order.
Parity is a $\mathbb{Z}_2$ symmetry with irreps $\sigma \in \{0, 1\}$; for index $i$, we write $p_i$ as the parity of $i$.
We will assume the fermionic tensors have well-defined total parity $0$ (even) or $1$ (odd).
Fermionic tensors may have additional symmetries, which may or may not be independent of parity. A common symmetry is $\mathbb{U}(1)$ symmetry, associated with fermion number or projected spin number. In this case, odd integer irreps of $\mathbb{U}(1)$ have parity $1$, while even integer irreps of $\mathbb{U}(1)$ have parity $0$. 



Every fermionic tensor is created with the legs in an  (arbitrary) specified order. In other words, when we write the tensor with indices, e.g.
$A_{ijkl}$
we mean that the indices have the definite order $i$ in the 0th position, $j$ in the 1st position, etc., which is not the same tensor as e.g. $A_{jikl}$.

Fermionic tensors support modified transposition and contraction operations that use the additional parity and index-order data.
Consider the tensor $\tilde{A}$ obtained by swapping two legs of $A$, i.e., transposing $i,j$. In a simple array $\tilde{A}_{ji\ldots} = A_{ij\ldots}$, but in a fermionic tensor, this transposition is defined with additional signs,
\begin{align}
    \tilde{A}_{ji \ldots} = (-1)^{p_ip_j} A_{ij \ldots} \label{eq:transposition}
\end{align}
Fermionic tensor contraction is similarly modified.
Consider the analog of Eq.~\ref{eq:simplecontract}
\begin{align}
    C_{ijlmnr} = \sum_k^F A_{ijkl} B_{mnkr}
\end{align}
where $A$, $B$ are fermionic tensors and the notation $\sum^F$ denotes a fermionic contraction.
Then, the above is evaluated as a standard contraction by first moving the indices that are to be contracted to be adjacent to each other, i.e.,
\begin{align}
    C_{ijlmnr} &= \sum_k \tilde{A}_{ijlk} \tilde{B}_{kmnr} \notag\\
    &= \sum_k (-1)^{p_kp_l}  {A}_{ijkl}
    (-1)^{p_k(p_m +p_n)}
    {B}_{mnkr} \notag \\
    &=
    \sum_k (-1)^{p_k(p_l+p_m+p_n)} A_{ijkl} B_{mnkr} \label{eq:standardcontract}
\end{align}
where the minus signs follow the rule for transposition in Eq.~(\ref{eq:transposition}), and the addition of parities follows $\mathbb{Z}_2$ arithmetic.

\begin{figure}
    \centering
    \includegraphics[width=1.0\columnwidth]{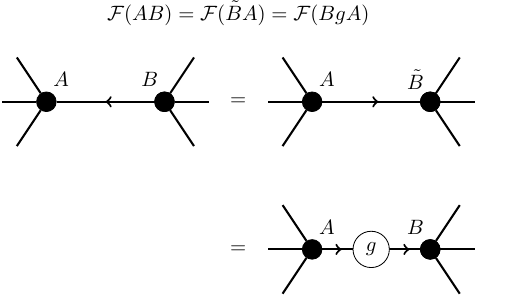}
    \caption{Swap rule for two fermionic tensors $A$ and $B$ for the fermionic contraction, $\mathcal{F}(AB)=\mathcal{F}(\tilde{B}A)=\mathcal{F}(BgA)$ (where we have identified $\tilde{A}=A$). A diagonal bond parity tensor $g$ is inserted between the tensors $A$ and $B$ after the swap. }
    \label{fig:swaprule}
\end{figure}

In a standard tensor network contraction, $\mathcal{C}(AB) = \mathcal{C}(BA)$. However, because of the additional parity signs, in general, the fermionic contraction $\mathcal{F} (AB)$ is not the same as $\mathcal{F}(BA)$. Instead,
\begin{align}
    \mathcal{F} (AB) = \mathcal{F}(\tilde{B}\tilde{A}) = \mathcal{F} (B g A) \label{eq:fermiswap}
\end{align}
where $g$ is a diagonal bond parity tensor defined below.
The above relation is known as the swap rule~\cite{pizorn2010fermionic}, and is illustrated in Fig.~\ref{fig:swaprule}. To derive it, consider the example
\begin{align}
    T_{mrks} &= \sum^F_n A_{mnr} B_{kns} \notag \\
    \tilde{T}_{ksmr} &= \sum^F_n \tilde{B}_{kns}  \tilde{A}_{mnr} = (-1)^{(p_m+p_r)(p_k+p_s)} T_{mrks} \label{eq:swapdefn}
\end{align}
We can then rewrite the fermionic contractions in Eq.~\eqref{eq:swapdefn} as standard contractions. Then
\begin{align}
    T_{mrks} &= \sum_n (-1)^{(p_r+p_k)p_n}  A_{mnr}B_{kns}\notag\\
    \tilde{T}_{ksmr} &= \sum_n (-1)^{(p_s+p_m)p_n}  \tilde{B}_{kns} \tilde{A}_{mnr} \notag \\
    &= \sum_n (-1)^{(p_m+p_r)(p_k+p_s)+(p_r+p_k)p_n}  A_{mnr}B_{kns}
    \label{eq:swapdefn2}
\end{align}
Following some $\mathbb{Z}_2$ additions we find $(p_m+p_r)(p_k+p_s)+(p_r+p_k)p_n + (p_s+p_m)p_n = p_A p_B + p_n$.
This allows us to identify
\begin{align}
    \tilde{A}_{mnr} &=  A_{mnr} \label{eq:swapeq} \notag\\
    \tilde{B}_{kns} &= (-1)^{p_n} (-1)^{p_Ap_B} B_{kns}
\end{align}
(where we have arbitrarily put all signs onto  $\tilde{B}$), or alternatively, define the bond parity tensor in Eq.~\ref{eq:fermiswap} as
\begin{align}
    g_n = (-1)^{p_n} (-1)^{p_{A}p_B}.
\end{align}

Fermionic tensors also support matrix decompositions such as QR and SVD. {The matrix decompositions are performed in the same way as the symmetric tensors after fermionic tensors are transposed according to ordering of the row and column indices, e.g., $A_{kijl} \rightarrow \Tilde{A}_{ijkl} \rightarrow \Tilde{A}_{ij,kl}$. }

\subsection{Fermionic tensor networks}

We now consider the fermionic tensor network defined by the contraction of many fermionic tensors. Here, one can use different but equivalent formalisms that we refer to as the ``globally ordered'' and ``locally ordered'' formalisms. In the globally ordered formalism, discussed clearly in Ref.~\cite{pizorn2010fermionic}, the fermionic tensors entering into the tensor network are assumed to be contracted sequentially from left to right. The tensors are then stored in the tensor network together with a global positional index. This has the advantage of familiarity for those new to tensor networks because the manipulations resemble those in second quantization, where operator expressions also involve fermionic operators in a definite order. However, the implementation needs to keep track of a global index for each tensor. This can lead to additional complexity in the implementation, as the contraction of two tensors is no longer a completely local operation but must refer to the other tensors in between.

In the locally ordered formalism, only the relative order of two tensors (an ``arrow'') with respect to the contraction of a shared fermionic bond is stored. If a tensor shares multiple bonds with another, the arrows need not point in the same direction on each bond 
.
This formalism is implicitly used in  Grassmann tensor networks~\cite{gu2010grassmann, gu2013grassmann, gu2013honeycomb} and is also the idea behind $\mathbb{Z}_2$ graded tensor networks~\cite{bultinck2017fermionic, bultinck2018fpeps, mortier2024fermionic} (also see \cite{dong2019fpeps, ZHANG2024109355}).
It is natural on arbitrary graphs as the ordering information can be stored only with the bonds, and contractions are truly pairwise operations. The globally ordered formalism can then be seen in the locally ordered language as working on a graph where all arrows form a directed acyclic graph.

We note that bosonic or spin degrees of freedom can be easily mixed into fermionic tensor networks by treating them as entirely even parity legs.

Our implementation of fermionic tensor networks for deterministic PEPS (used, e.g., in Ref.~\cite{Lee2023}) originally used the globally ordered formalism, while our variational Monte Carlo fermionic PEPS calculations used the locally ordered formalism~\cite{Lee2023, liu2024tensor}. Both conventions are supported in \texttt{quimb}, but as the local ordering is most natural for arbitrary graphs, as clearly pointed out in Ref.~\cite{gu2013grassmann, dong2019fpeps, mortier2024fermionic}, this is now the default fermionic formalism. Below, however, we start for pedagogical reasons with the globally ordered formalism, then describe its relation to the locally ordered formalism.

\subsection{Globally ordered formalism}

\begin{figure}
    \centering
    \includegraphics[width=1.0\columnwidth]{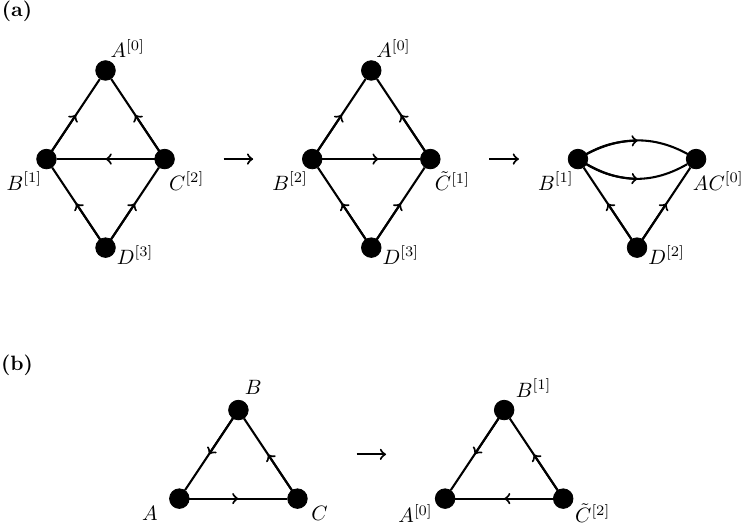}
    \caption{(a) Given the ordered fermionic tensor network, $\mathcal{T}=\mathcal{F}(A^{[0]}B^{[1]}C^{[2]}D^{[3]})$ on the graph, the relative ordering in the indexes determines the arrow directions in the bonds. The graph with arrows forms a directed acyclic graph (DAG). When contracting $A^{[0]}$ and $C^{[2]}$, the relative ordering between $B^{[1]}$ and $C^{[2]}$ needs to be swapped to be $\Tilde{C}^{[1]}$ and $B^{[2]}$. During the swap, the direction of the arrow between $B^{[1]}$ and $C^{[2]}$ is converted. (b) When the graph has a cycle, the global ordering of the tensors cannot be assigned. Nonetheless, any directed graph with cycles can be made into a DAG by converting some arrows to assign the order of the tensors.   }
    \label{fig:ftndag}
\end{figure}

As indicated in the swap rule, unlike in a standard tensor network, the
order of contraction of the fermionic tensors matters in defining the value of $\mathcal{T}$. In the globally ordered formalism, we define $\mathcal{T}$ through the sequence
\begin{align}
    \mathcal{T} = \mathcal{F} (A^{[0]} B^{[1]} C^{[2]} \ldots) \label{eq:fermionictn}
\end{align}
where we store the additional data on the initialization order in the form of an integer position for each fermionic tensor (the superscripts in Eq.~\ref{eq:fermionictn}). $\mathcal{T}$ is then obtained by contracting the network in sequence from left to right. Transposition of $\mathcal{T}$ (e.g., when taking the hermitian conjugate to go from $|\Psi\rangle$ to $\langle \Psi|$) involves reversing the order of the tensors appearing in Eq.~\ref{eq:fermionictn}, as well and reversing the indices of all the tensors.

To contract $\mathcal{T}$ in a different contraction order, we manipulate the positional index of the tensors using the swap rule. For example, in the tensor network
\begin{align}
\mathcal{T} = \mathcal{F}(A^{[0]}B^{[1]}C^{[2]}D^{[3]})
\end{align}
illustrated in Fig.~\ref{fig:ftndag}a, we may wish to first contract tensors $A^{[0]}$, $C^{[2]}$, which are not adjacent in the positional order, so we first move $C$ to be adjacent to $A$,
\begin{align}
\mathcal{T} = \mathcal{F}(A^{[0]}\tilde{C}^{[1]}B^{[2]}D^{[3]})
\end{align}
where we have accumulated bond parity tensors onto $\tilde{C}$. Note that one obtains a non-trivial bond parity tensor every time we move $C$ past a tensor with which it shares a contracted bond.
In addition, swapping odd parity tensors past each other, even when they are not connected by a bond, generates a simple global phase.

The global ordering of tensors may seem to be a separate concept from the graph structure of the TN. However, it is, in fact, simply related. The integer ordering can be used to define an arrow annotation of the bonds of the tensor network such that the bonds form a directed acyclic graph (DAG) (see Fig.~\ref{fig:ftndag}a).
In the DAG, every bond arrow is consistent with the relative ordering of the index positions between two tensors, i.e., it connects a tensor that is earlier in the positional order with a tensor that is later in the positional order. Working with global positions is thus equivalent to working with a DAG.

In this DAG picture, we can also visualize what happens when one changes the global ordering.
When moving tensors past each other to construct a new global order, one needs to
construct a new DAG corresponding to the new order (see sequence in Fig.~\ref{fig:ftndag}a). In general, this requires the reversal of multiple arrows in the original DAG, and these reversals correspond to the bond parity tensors that are discussed in Sec.~\ref{sec:fermionicten}.

\subsection{Locally ordered formalism}

From here, it is a short leap to the locally ordered formalism. It is clear from the above that it is not necessary to limit ourselves to DAGs, but one can use other arrow graphs, where the arrows only define the local ordering of two tensors in a given contraction. In this case, the value of the fermionic tensor network is given by
\begin{align}
    \mathcal{T} = \mathcal{G} (A B C \ldots) \label{eq:fermionictn2}
\end{align}
where $\mathcal{G}$ denotes the fermionic contraction following the additional data where every contracted bond has an arrow, and there is no longer any global ordering of the tensors. The computation of the bond parity is needed only on the bond between two tensors when they are contracted: if they are contracted in the direction of the arrow that bonds them, there is no parity, while if they are contracted in the opposite order, we compute the bond parity (or equivalently, first reverse the arrow).
The above assumes that we are working with even parity tensors, so there is no global phase. However, if there are odd total parity tensors in $\mathcal{T}$, we also need to track the phase associated with the ordering of the odd parity tensors.
It is convenient to do so by implementing odd parity tensors using an additional dummy (size 1) odd index to create an even parity tensor.
The dummy indices are then propagated to the final tensor that is the result of the full contraction of the tensor network, and sorting the remaining dummy indices according to a chosen global ordering convention yields the phase defined by the odd parity tensors (according to the same convention).

Clearly, not every locally ordered choice of arrows is consistent with a global ordering; for example, if the arrows form a cycle, then it is not a DAG. However, we can reverse the arrows, which is equivalent to placing a parity tensor on the reversed bond (Fig.~\ref{fig:ftndag}b). If we do so to obtain a DAG of fermionic tensors (with parity tensors absorbed), this may be re-expressed in terms of the globally ordered contraction operation $\mathcal{F}$.



\section{Hyperoptimized fermionic contraction}
\label{sec:hyper}

The contraction algebras above, either in the globally ordered or locally ordered conventions, allow us to choose an arbitrary contraction order between the tensors by appropriately applying the swap rule. Thus, we are free to choose a sequence of contractions optimized for the computational costs.

For an exact contraction, we can easily estimate the theoretical compute and memory costs for a given contraction order. Unfortunately, the space of contraction trees grows too fast with the number of tensors for a brute force optimization of contraction cost to be possible.
Ref.~\cite{Gray2021hyperoptimized} introduced a heuristic to search over contraction trees of standard tensor networks defined on arbitrary graphs by generating contraction trees using a graph partitioning technique.  The space of contraction trees searched over was controlled by a set of hyperparameters that could be optimized, leading this to be termed a hyperoptimized contraction.

In the case of approximate contraction, further considerations are necessary to achieve an optimized cost because QR and SVDs (i.e., performing tensor compression) need to be considered in the contraction tree, and the error of the approximate contraction is also a relevant metric.
Ref.~\cite{Gray2024hyperoptimized} extended the hyperoptimization technique to approximate contraction trees of standard tensor networks through the following observations: (i) minimizing contraction cost (either memory or computational time) is generally a good proxy for optimizing the error of the approximate contraction because a good contraction tree allows for larger compressed bond dimension $\chi$ for a given cost, and (ii) there are some reasonable heuristics to determine when to perform compression: either as early as possible (once a large bond dimension appears, which is a good heuristic when optimizing cost for a given accuracy), or as late as possible (i.e., large bonds are only compressed when the tensor on which they appear is about to be contracted with another one), which is a good heuristic when optimizing the accuracy for a given $\chi$, i.e., the size of the largest tensor operation. Given a choice of early or late compression, the optimization of approximate contraction trees is then tied to optimizing only the contraction order itself, and a similar hyperoptimization strategy to the one used for exact contraction can be used. In particular, Ref.~\cite{Gray2024hyperoptimized} introduced several families of contraction tree generators, each with hyperparameters, which could then be optimized to minimize the memory or computational cost.

For fermionic tensor networks, changing the contraction order requires computing additional parities, but this does not incur any appreciable computational cost. Thus, determination of an optimized contraction order can take place from the tensor network graph $\mathcal{T}$ without any need to consider the fermionic nature of the tensors.

\section{Implementation}

\label{sec:implementation}

We have implemented the above fermionic tensor network algebra
in an Abelian block-sparse and fermionic array library \texttt{symmray}~\cite{symmray_code}, which acts as a new backend for \texttt{quimb}~\cite{gray2018quimb}. 
No significant changes are required, and our fermionic implementation can thus take advantage of the general tensor network infrastructure in \texttt{quimb}, including its ability to perform hyperoptimized exact and approximate contraction.
Below, we discuss some more specific implementation details.

\subsection{Tensor operations and index fusion}

From a function perspective, the main changes are that we must redefine unary and binary tensor operations, such as transposition and contraction, to incorporate the new bond parity contributions. If using a global ordering throughout the calculation, we introduce additional bookkeeping to track the global position of the tensors in the total contraction $\mathcal{F}$. 

In addition, we also implement fusion (combining multiple indices into one, either for efficient contraction or linear algebra decompositions), which involves multiple parity contributions and is worth a small discussion. The first arises from the fermionic transposition to make the indices adjacent. Second, all arrows in the fusion set are required to point in the same direction (as the first arrow), which may require flipping arrows. Third, for a fused bond with an inward arrow, we reverse the ordering of the fermionic modes so we obtain the same parity contributions as if the modes were contracted one by one, i.e. $\sum_{ijk} A_{m(ijk)} B_{(ijk)n} = \sum_{i}\sum_{j}\sum_{k} {A}_{m(ijk)} \tilde{B}_{(kji)n}$, where $(\ldots)$ denotes the fused index, and $\tilde{B}$ indicates that the parity contributions of $(ijk)\to (kji)$ have been accumulated into tensor $B$.


\subsection{Tensor network initialization}

When initializing an arbitrary tensor network, the direction of the arrows can be chosen completely arbitrarily (locally ordered formalism) or they can be inferred from a given global ordering of the tensors (globally ordered formalism). In practice, by default, we initialize the arrows so that they point from later to earlier sites, which is consistent with a global ordering, and a DAG. In principle, there is an additional choice of the incoming/outcoming symmetry label of each tensor leg (which must also be done consistently), but we can choose this labelling to align with the arrow directions (i.e. the arrow points from $+$ to $-$), and this is done by default. 

\begin{figure}
    \centering
    \includegraphics[width=1.0\columnwidth]{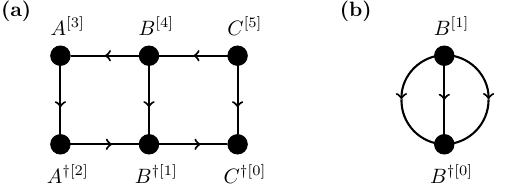}
    \caption{(a) Norm tensor network for $| \Psi \rangle = \mathcal{F}(A^{[0]}B^{[1]}C^{[2]}) $ as $\langle \Psi | \Psi\rangle = \mathcal{F}(C^{\dag[0]} B^{\dag[1]} A^{\dag[2]} A^{[3]} B^{[4]} C^{[5]})$. (b) Norm for the partial `cluster' formed by taking only the central site $\mathcal{F}(B^{\dag[0]} B^{[1]})$, requiring some change of arrow directions.}
    \label{fig:norm}
\end{figure}

As an example of this initialization, we consider the case of setting up the tensor network corresponding to the norm $\langle \Psi | \Psi\rangle$. For  a $|\Psi\rangle$ given by $\mathcal{F}(A^{[0]}B^{[1]}C^{[2]})$, we can obtain $\langle \Psi|$ by transposition, giving the overall globally ordered tensor network $\mathcal{F}(C^{\dag[0]} B^{\dag[1]} A^{\dag[2]} A^{[3]} B^{[4]} C^{[5]})$. Note that the tensors in the bra all occur before those in the ket, and inferring the arrow directions of the DAG from the global ordering yields the picture shown in Fig.~\ref{fig:norm}(a), where the bra arrows are all the reverse of those in the ket.

As a second example, in the cluster algorithm implemented later in Sec.~\ref{sec:examples}, we extract part of the tensor network (a cluster) in the bra and contract it with the same cluster in the ket; e.g. consider extracting only the $B$ tensors from the tensor network expression above. The norm then becomes the tensor network $\mathcal{F}(B^{\dag[0]} B^{[1]})$, corresponding to the DAG in 
Fig.~\ref{fig:norm}(b). Note that the arrows joining the $B$ tensors in Fig.~\ref{fig:norm}(b), as inferred from the DAG, are different to the arrows joining the $B$ tensors in Fig.~\ref{fig:norm}(a), as some are reversed. This is equivalent to implementing the bond parity tensor coming from ``partial trace'' operations discussed in Ref.~\cite{mortier2024fermionic}.

\section{Examples}

\label{sec:examples}
To illustrate the application of our fermionic library, we consider fermionic PEPS simulations of Hubbard models on finite graphs with complicated geometries: the 3D diamond lattice Hubbard model and random regular graph Hubbard models of degree 3. The Hubbard model Hamiltonian
can be expressed as:
\begin{equation}\label{eq:fermi-hubbard-ham}
	H = -t\sum_{\langle i,j \rangle \sigma}\left(a_{i\sigma}^{\dagger}a_{j\sigma} + \mathrm{h.c.}\right) + U\sum_{i}n_{i\uparrow}n_{i\downarrow} ,
\end{equation}
where $\langle i,j \rangle$ denotes nearest neighbors within the lattice and $U$ is the on-site
repulsion. In terms of the Hubbard sites, the Hamiltonian is geometrically 2-local, i.e., it can be written as $H = \sum_t h_t$, where $h_t$ acts only two adjacent sites. We will work in energy units of $t$.

\begin{figure}[t]
\centering
\includegraphics[width=0.75\columnwidth]{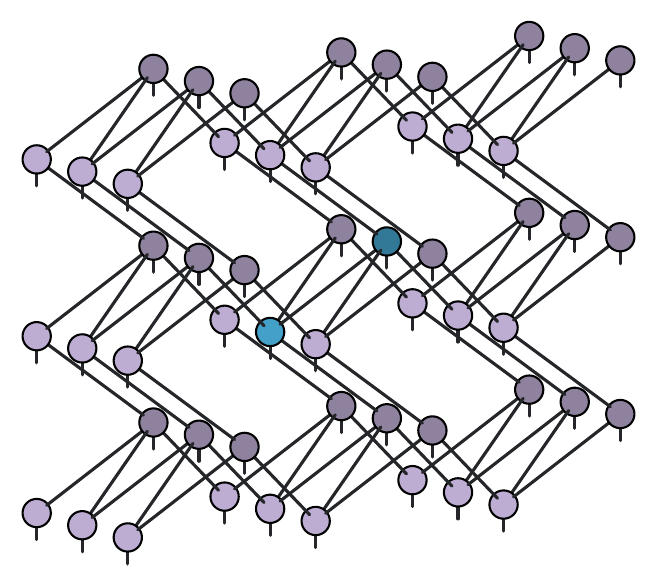}
\caption{Geometry of a $3 \times 3 \times 3$ diamond lattice. The central sites in blue correspond to the primitive cell. }\label{fig:ftn_dmd}
\index{figures}
\end{figure}

\subsection{3D diamond lattice Hubbard model}

We consider the half-filled Hubbard model on the 3D diamond lattice.
We approximate the ground-state by a PEPS tensor network with $\mathbb{U}(1)$ symmetry and optimize the ground-state using simple update (SU) imaginary time evolution~\cite{jiang2008accurate} as implemented in \texttt{quimb}. The simulation is carried out at a fixed total particle number using  $\mathbb{U}(1)$ symmetric tensors.
The lattice is bipartite, so the auxiliary field quantum Monte Carlo (AFQMC) does not suffer from a sign problem and is exact up to statistical error. We use \texttt{Dice}~\cite{dice_code} to compute the AFQMC ground-state energy as a benchmark.

The specific diamond structures are constructed as $(L_x, L_y, L_z)$ finite lattices of primitive diamond unit cells, where $L_x, L_y, L_z$ are integers, denoting multiples of the 3 primitive lattice vectors. An example of a $3 \times 3 \times 3$ diamond lattice is shown in Fig.~\ref{fig:ftn_dmd}. Since the primitive cell has two sites, the total number of Hubbard sites is $2 \times L_x \times L_y \times L_z$. We consider calculations on $3 \times 3 \times 3$, $4 \times 4 \times 4$, and $5 \times 5 \times 5$ diamond graphs. Thus, the largest lattice has 250 Hubbard sites.

To evaluate the energy of the PEPS, we considered two strategies. The first is to use a direct approximate contraction $\langle \Psi | H | \Psi\rangle$ (we contract separately for each term in $H$) for the entire finite diamond graph using the hyperoptimized approximate contraction strategy in Sec.~\ref{sec:hyper}. The second is to use a cluster approximation, detailed below.

\begin{figure}[t]
    \centering
    \includegraphics[width=1.0\columnwidth]{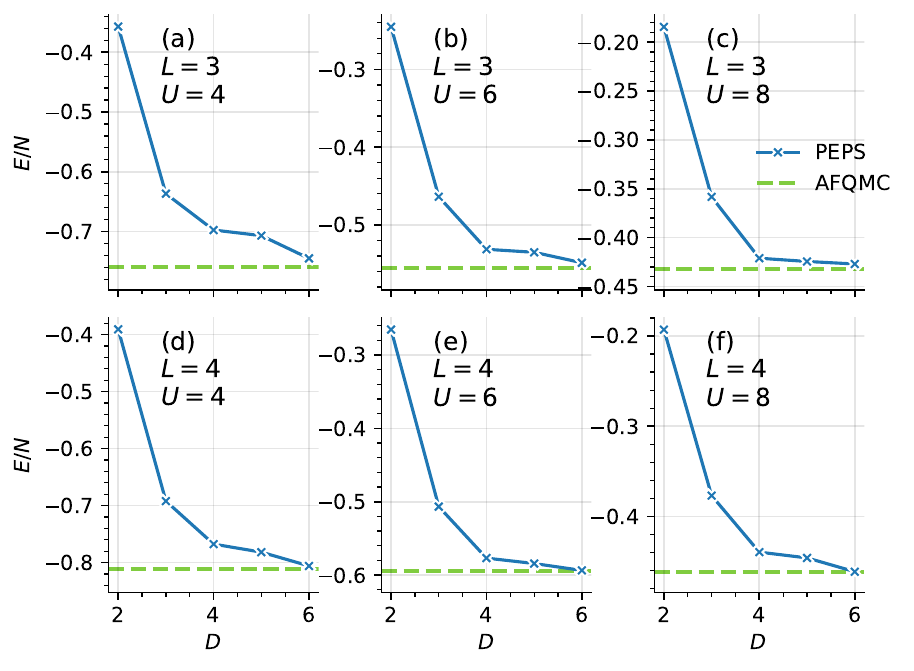}
    \caption{Ground-state energies for the Hubbard model on the $3 \times 3 \times 3$ (top row) and $4 \times 4 \times 4$ (bottom row) diamond lattice with $U=4, 6, 8$.
    The PEPS results, which are optimized with SU, and with the energy computed using full approximate contraction, are compared with AFQMC.}\label{fig:ftn_dmd34}
    \index{figures}
    \end{figure}

We first show results for the $3\times 3\times 3$ and $4 \times 4\times 4$ diamond lattices using energies evaluated by hyperoptimized approximate contraction. We optimize a PEPS ground-state with $D=6$ and for the approximate contraction hyperoptimized path we use the  default search parameters in \texttt{quimb} with $\chi=32$ and $\chi=16$ for the two lattice sizes respectively. We compare against exact results from AFQMC in Fig.~\ref{fig:ftn_dmd34} and in Table~\ref{tab:ftn_dmd}.
At the $D=6$ level, the PEPS energies are converged to within 1-2\% of the exact AFQMC benchmark.
From Fig.~\ref{fig:ftn_dmd34}, we see that the PEPS energies are converging quite rapidly with $D$. The results are most accurate for larger $U$. This reflects the relative difficulty PEPS has in simulating more itinerant fermions at smaller $U$.

\begin{figure}[t]
    \centering
    \includegraphics[width=1\columnwidth]{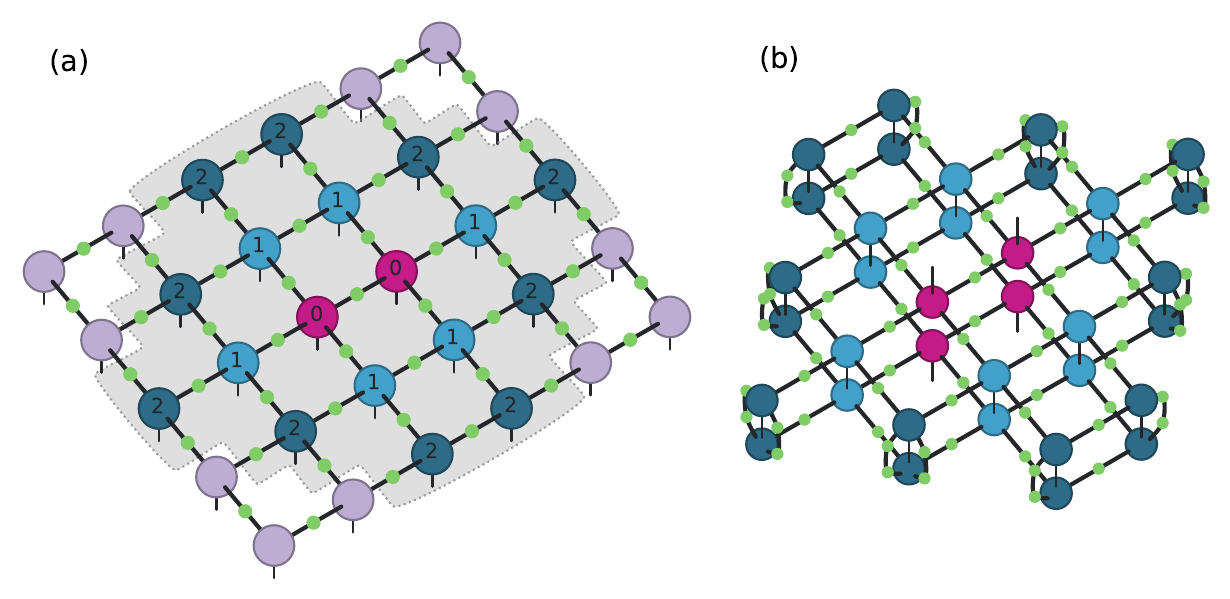}
    \caption{
    Schematic diagrams for the cluster approximation illustrated for a radius $r=2$ on a square lattice.
    (a) Full wavefunction with simple update gauges (green dots) associated with each bond. The shaded region includes the target two sites (pink, $r=0$), their neighbors (light blue, $r=1$) and next nearest neighbors (dark blue, $r=2$) as well as the gauges associated with any bonds crossing beyond this region, which approximate the contribution from remaining tensors (light purple).
    (b) The approximate reduced density matrix formed by taking this shaded region and partially contracting every bond other than the physical indices of the target sites.
    }\label{fig:ftn_clusterc}
    \index{figures}
\end{figure}

We next compute the energy by a cluster approximation, as used, e.g., in Refs.~\cite{picotNematicSupernematicPhases2015,picotSpin$S$KagomeQuantum2016,jahromiUniversalTensornetworkAlgorithm2019,jahromiThermalBosons3d2020,vlaarSimulationThreedimensionalQuantum2021}. In this case, for each local term $h_t$ in $H$, we define a cluster of tensors around the pair of sites on which $h_t$ acts out to a finite radius (graph distance) $r$. (We show a 2D example of this in Fig.~\ref{fig:ftn_clusterc}). This approximation is motivated by the fact that contracting the environment of $h_t$ is the main cost in contracting the tensor network, but in a tensor network without loops that are appropriately gauged, this environment contraction just yields the identity. Thus, here,
the environment tensors outside of the cluster are approximated by the simple update gauges on each of the dangling bonds. The contraction of the cluster can take advantage of the approximate contraction to reduce the cost further. In this work, we use $\chi = 32$ for the approximate contraction of the cluster.

\begin{figure}[t]
\centering
\includegraphics[width=1.0\columnwidth]{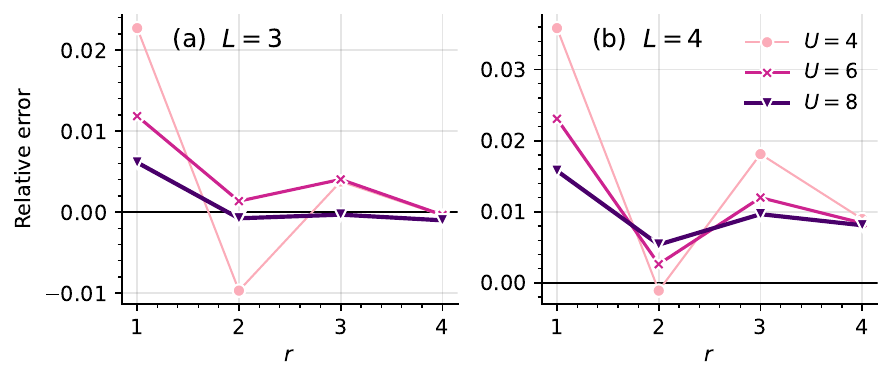}
\caption{
The relative difference between PEPS energy cluster approximation and full approximate contraction as a function of cluster radius $r$ for systems sizes (a) $L=3$ and (b) $L=4$ and varying $U$ at $D=6$.
}\label{fig:ftn_cluster_diff}
\index{figures}
\end{figure}

The smaller lattices provide an opportunity for us to assess the cluster approximation for energy evaluation. In Fig.~\ref{fig:ftn_cluster_diff}, we assess it as a function of radius $r$, comparing to the results from the full approximate contraction (with $\chi=32$ for the $3\times3\times3$ lattice and $\chi=16$ for the $4\times 4\times 4$ lattice).
For the $3\times3\times3$ case, the cluster approximation achieves a relative error of less than 1\% for $r\geq 2$; the errors are smaller for larger $U$, where the correlations are more local. For the $4 \times4\times4$ structure, however, the smallest difference between the cluster approximation and the full approximate contraction actually occurs at $r=2$.
Since the $4\times 4\times 4$ approximate contraction is itself not very accurate (as $\chi=16$ is quite small), this does not mean that the $r=2$ cluster gives the best result. Overall, however, the above suggests that the cluster approximation error with $r=2$ or $r=3$ is comparable to the accuracy of the ansatz itself for $D=6$, with the largest error appearing for $U=4$, the most itinerant system.

\begin{figure}[t]
\centering
\includegraphics[width=1.0\columnwidth]{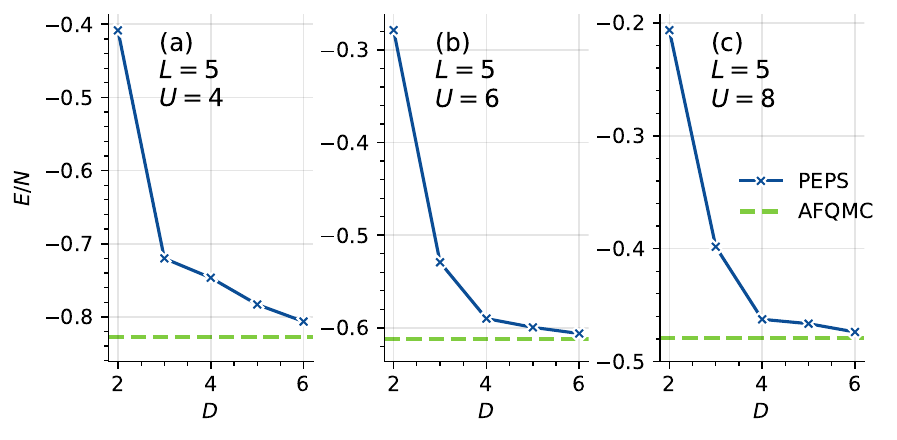}
\caption{Ground-state energies of the Hubbard model on the $5 \times 5 \times 5$ diamond lattice with $U=4, 6, 8$.
The PEPS results, which are optimized with SU, with the final energy computed with the $r=3$ cluster approximation, are compared with AFQMC.
}\label{fig:ftn_dmd_L5}
\index{figures}
\end{figure}

In Fig.~\ref{fig:ftn_dmd_L5} and Table~\ref{tab:ftn_dmd}, we show the SU optimized PEPS ground-state energies with bond dimension up to $D=6$ on the $5 \times 5 \times 5$ lattice where the final energy is evaluated using the $r=3$ cluster approximation.
We see a rapid decrease of the energy with $D$, similar to in the smaller lattices. The relative error compared to AFQMC is seen to be similar to the smaller lattice sizes, with the exception of $U=4$, where we have a slightly larger error of $\sim 3\%$.

\begin{table}[ht!]
\centering
\begin{tabular}{cccc}
\hline
\hline
\textrm{Size}&
\textrm{$U$}&
\textrm{SU ($D=6$)}&
\textrm{AFQMC}\\\\
\hline
       & 4  & -0.745, -0.745  & -0.760\\
$3 \times 3 \times 3$  & 6  & -0.549, -0.549 & -0.556\\
       & 8  & -0.427, -0.427 & -0.432\\
\hline
       & 4  & -0.806, -0.799 & -0.812 \\
$4 \times 4 \times 4$  & 6  & -0.594, -0.589 & -0.594 \\
       & 8  & -0.461, -0.457 &  -0.461 \\
\hline
       & 4  & NA, -0.806 & -0.828 \\
$5 \times 5 \times 5$  & 6  & NA, -0.606 & -0.612 \\
       & 8  & NA, -0.474   & -0.479 \\
\hline
\end{tabular}
\caption{\label{tab:ftn_dmd}
    Ground-state energies of the Hubbard model on diamond lattices. The first entry in the SU column is computed from approximate contraction of the whole lattice and the second from the cluster approximation with $r=3$.}
\index{tables}
\end{table}

\begin{figure}[t]
    \centering
    \includegraphics[width=1.0\linewidth]{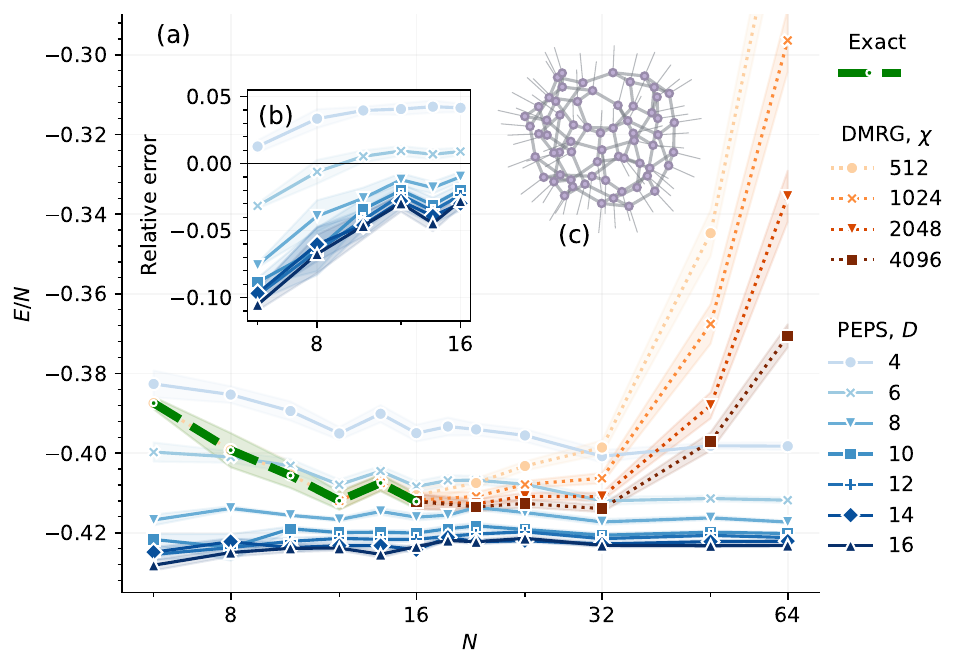}
    \caption{
    Fermionic PEPS applied to the Fermi-Hubbard model on random 3-regular graphs with $U=8$ at half-filling.
    The PEPS is optimized with SU, with energy computed using the $r=0$ cluster approximation.
    All lines show mean behavior across 20 instances with shaded areas denoting error on the mean.
    (a) Energy per site compared to exact diagonalization and DMRG as a function of number of sites, $N$, for varying PEPS dimension, $D$, and MPS bond dimension, $\chi$.
    (b) Error in energy compared to exact diagonalization.
    (c) Example PEPS geometry for a single instance at $N=64$.
    }
    \label{fig:rrg-fh}
\end{figure}

\subsection{Random regular graphs}

We next move on to the more unusual geometry of random 3-regular graphs, which are neither bipartite nor short-range and thus quite challenging for existing methods.
We stick with the Hubbard model given by Eq.~\ref{eq:fermi-hubbard-ham} in the strongly interacting regime of $U=8$ at half filling.
Given the complex geometry, we again choose to evaluate the energy using the cluster approximation, which is generally non-variational but provides a useful overview of trends.
We do know as the number of sites $N\rightarrow\infty$ that this approximation should become exact however, given the graph loops become infinitely long, {and the local geometry then becomes tree-like.}
We take 20 instances for each graph size $N$ and optimize fermionic PEPS with a variety of bond dimensions, $D$, using the simple update.
The energy is then evaluated using the simple update gauges at cluster radius $r=0$.
Up to $N=16$ we can compare to exact diagonalization, and beyond that we also compare to DMRG for a range of maximum bond dimensions $\chi$.
We compute the DMRG results using the \textsc{block2} package\cite{zhai2021low,Zhai2023}, without spin adaptation.
We converge each bond dimension $\chi$, then use that state as the initial guess for the next, doubled $\chi$.
At $N{=}64$ and $\chi{=}4096$ this requires multiple days of compute on 16 cores, compared to about ${\sim}15$ minutes for the $D{=}16$ PEPS computation.

In Fig.~\ref{fig:rrg-fh}(a), we show the results for the energy per site.
Across the board, we see convergence in $D$ for the PEPS results.
At the smallest sizes (with the highest density of short loops), the PEPS energy evaluations are about 10\% lower than the exact results.
This drops to about 2\% at $N=16$.
The error is shown in the inset Fig.~\ref{fig:rrg-fh}(b), where the expected decreasing trend is clear.
At these small sizes, exact contraction of the PEPS energy (not shown) gives closer results -- around 1\% error -- suggesting the cluster approximation is the major source of error, rather than the ansatz and optimization itself.
Beyond $N=16$, we see the PEPS energy per site starts to stabilize towards the presumed infinite graph limit.
However, the DMRG results, which are very accurate for small sizes, start to diverge for different $\chi$ at this point.
Although the sites are ordered to minimize long range entanglement in DMRG, eventually the amount of long-range entanglement in the 1D ordering becomes too challenging for the accessible $\chi$,
and in this regime fermionic PEPS shows a distinct advantage.
An example PEPS for such a geometry at $N=64$ is depicted in Fig.~\ref{fig:rrg-fh}(c).

\section{Conclusion}

We have described our implementation of a numerical framework for tensor network simulations of fermions on arbitrary geometry lattices, with a discussion of the different conventions that can be used to incorporate the fermionic signs. The compatibility of this framework with the existing technology for hyperoptimized tensor network contraction allows for its immediate deployment in efficient simulations on complex graphs, as we showed with simple benchmarks on the 3D Hubbard model on the diamond lattice, and 3-regular random graphs.
The implementation in \texttt{quimb} provides a starting point for fermionic quantum many-body simulations in a variety of applications in condensed matter theory, quantum information, and quantum chemistry.

\begin{acknowledgments}
    The initial development of this library by YG and HZ that used the \texttt{pyblock3} fermionic sparse tensor library~\cite{Zhai2023,pyblock3}, was supported by the US National Science Foundation through grant no. CHE-2102505. The subsequent \texttt{symmray} implementation by JG (now the default implementation in \texttt{quimb}) was supported by the Quantum Utility through Advanced Quantum Computational Algorithms Center, funded by the US Department of Energy, Office of Science, through grant no. DE-SC0025572.
\end{acknowledgments}

\bibliography{main}

\end{document}